\documentclass[aps,article]{revtex4}
\usepackage{amsmath,amssymb,amsfonts,graphicx,hyperref}
\newcommand{\beq}{\begin{equation}}
\newcommand{\eeq}{\end{equation}}
\newcommand{\bea}{\begin{eqnarray}}
\newcommand{\eea}{\end{eqnarray}}

\begin{document}


\title{Vacuum polarization at the boundary of a topological insulator}

\author{C. R. Muniz}
\email{celio.muniz@uece.br}\affiliation{Grupo de F\'isica Te\'orica (GFT), Universidade Estadual do Cear\'a, Faculdade de Educa\c c\~ao, Ci\^encias e Letras de Iguatu, CEP 63500-000, Iguatu, Cear\'a, Brazil.}

\author{M. O. Tahim}
\email{makarius.tahim@uece.br}\affiliation{Grupo de F\'isica Te\'orica (GFT), Universidade Estadual do Cear\'a, Faculdade de Educa\c c\~ao, Ci\^encias e Letras do Sert\~ao Central, CEP 63900-000, Quixad\'a, Cear\'a, Brazil.}

\author{G. D. Saraiva}
\email{gilberto.saraiva@uece.br}\affiliation{Grupo de F\'isica Te\'orica (GFT), Universidade Estadual do Cear\'a, Faculdade de Educa\c c\~ao, Ci\^encias e Letras do Sert\~ao Central, CEP 63900-000, Quixad\'a, Cear\'a, Brazil.}

\author{M. S. Cunha }
\email{marcony.cunha@uece.br}\affiliation{Grupo de F\'isica Te\'orica (GFT), Centro de Ci\^encias e Tecnologia, Universidade Estadual do Cear\'a, CEP 60714-903, Fortaleza, Cear\'a, Brazil.}

\begin{abstract}

 In this paper we study the polarized vacuum energy on the conducting surface of a topological insulator characterized by both $Z_2$ topological index and time reversal symmetry. This boundary is subject to the action of a static and spatially homogeneous magnetic field perpendicular to it as well as of an electric field that is uniform near the considered surface and produced by a biased voltage, at zero temperature. To do this, we consider modifications in the Gauss law that arise due to the nonzero gradient of the axionlike pseudoscalar factor coupled to the applied magnetic field, which accounts for the topological properties of the system. Such a term allows us to find a correction to the induced charges which modifies the quantum vacuum of the spinor field regarding an ordinary surface. The polarized vacuum energy is calculated in both the weak-field approximation and in the general case, and since the found energy depends on a length defined on the boundary, we show that there is a radial density of force or a surface shear stress that tends to shrink it.

 \vspace{0.75cm}
\noindent{Key words: Vacuum polarization, Axion electrodynamics, Topological insulator.}
\end{abstract}

\maketitle

\section{Introduction}

Topological insulators (TIs) are electronic systems with a nonzero energy gap to the excited (conducting) states in the bulk, nevertheless possessing a gapless mode on their boundary surfaces (edges) \cite{Hasan}. The meaning of the term ``topological'' comes from the fact that the wave functions describing their electronic states span a Hilbert space with a nontrivial topology \cite{ryu}. Such systems are also characterized by a topological number that reveals how robust they are against external perturbations, including those ones related to the quantum vacuum.

A lot of theoretical work has been done since the original formulation by Casimir \cite{Casimir} concerning the macroscopic effect which is caused by modifications in the fluctuations of the quantum vacuum energy of the electromagnetic field and that bears his name. The effect originally described consists of an attraction force that arises between two ideal conducting plates closely spaced, parallel and uncharged, placed in a perfect vacuum at zero temperature. This phenomenon was experimentally confirmed ten years later with some level of precision. Numerous papers published since then have addressed various types of geometrical settings, quantum fields, materials, space-times, topologies and thermal conditions. These and the latest developments both theoretical and experimental, are described in detail in \cite{Bordag,Miloni}.

Effects related to the quantum vacuum can also be present in empty spaces with non-Euclidean topology including cosmological models, in general relativity or alternative theories of gravity \cite{Celio1,Celio2}. In these space-times there are no material borders, but the identification of conditions imposed on the quantum fields can play the same role as the one presented by these boundaries. When macroscopic fields are present, possible physical effects can arise due to the polarization of the quantum vacuum induced by these fields. Thus, it is interesting to verify how such effects operate over systems as topological insulators, in which the nontrivial topology is in their spaces of quantum states instead of in their physical spaces. In this investigation line, the literature presents some works \cite{Adolfo,Rodriguez,Alberto}.

In the present work we will study the polarized quantum vacuum associated to 2D charged fermions when an uniform electric field generated by a biased voltage and an equally uniform magnetic field act on the surface of a special class of TI's, labeled by a topological number the $Z_2$ index, which expresses the number of times that a 1D edge electronic state crosses the Fermi level between 0 and $\pi/a$, where $a$ is the lattice parameter, besides exhibiting time reversal (TR) symmetry (e.g., Bi$_2$Se$_3$). In these systems there is a term in the electromagnetic Lagrangian proportional to ${\bf \theta E\cdot B}$, derived from the axion electrodynamics \cite{wilck}, which, inside the TI, does not modify the usual Maxwell's laws valid in material media by virtue of the $\theta$ factor being a constant, despite changing the electromagnetic constitutive relations. This leads, for example, to the appearance of magnetization (polarization) induced by the electric (magnetic) field \cite{ando}. It is interesting to point out that, according to Ref. \cite{Li}, it is possible to study the breakdown of TR symmetry via magnetic fluctuations in order to $\theta$ acquire dynamics, becoming a field variable.

Although $\theta$ in our approach is not a dynamical field, on the TI conducting surface the Maxwell's laws are modified since ${\bf\nabla}\theta\neq0$. Thus, we will consider the variation in the Gauss' law due to the nonzero gradient of $\theta$ which is coupled to a static and spatially homogeneous magnetic field perpendicular to the TI boundary in order to find an effective electric charge that modifies the polarized vacuum energy on that edge, firstly in the weak-field limit, where the energy gap is approximately zero. Next, we will analyze the more general situation by considering the nonzero gap opened through incidence of stronger magnetic fields. In this case, we will use the result found in \cite{Visser} in order to analyze the polarized quantum vacuum of a spinor field in 1+2 dimensions. Since this new vacuum energy depends on a length defined on the boundary, there is a tangential force or a shear stress associated to it.

The present paper unites frontier areas of research, namely polarized vacuum effects and topological insulators, investigating a phenomenon that can be relevant in the technological applications of TIs, since the work predicts the occurrence of a dynamical effect due to the quantum vacuum which is possibly non-negligible.

The paper is organized as follows: In Sec. II, we present the electrodynamics of a TI with $Z_2$ topology and TR symmetry, obtaining a modification in the electric charges on its conducting boundary. In Sec. III, we reobtain the polarized vacuum energy density of a spinor field endowed with the found effective charge and interacting with a constant magnetic field on that boundary, in both the weak-field regimen and general case. Finally, in Sec. IV, we discuss the results.

\section{Correcting the Induced Electric Charge}

In this section, we will study how variations in the electrodynamics laws induced by the topology of TIs modify in turn the electric charge of the carriers. The electromagnetic Lagrangian density in a $Z_2$-type TI with TR symmetry can be written as \cite{wilck}
\begin{equation}\label{1}
\mathcal{L}_E=\frac{1}{8\pi}\left({\bf E}^2-{\bf B}^2\right)+\frac{c\alpha }{4\pi^2}\theta{\bf E}\cdot{\bf B},
\end{equation}
without source terms. The factor $\alpha=e^2/\hbar c$ is the fine structure constant. The second term of Eq. (\ref{1}) - the $\theta$-term, which encodes the topological features of TIs, is quite similar to the one which occurs in the axion electrodynamics, where $\theta(x)$ is a pseudoscalar field, for which its transformation in a temporal reversion, given by $t\rightarrow -t$, is $\theta(x)\rightarrow -\theta(x)$, and whose coupling with the electromagnetic field generates unusual effects as axion-photon oscillations \cite{raffelt} and shining light through walls \cite{Javier}.

The presence of the $\theta$-term changes the electromagnetic constitutive relations, in the form
\begin{eqnarray}\label{2}
{\bf D}={\bf E}-\frac{\theta c\alpha}{\pi}{\bf B}\\
{\bf H}={\bf B}+\frac{\theta c\alpha}{\pi}{\bf E},
\end{eqnarray}
in a such way that within the TI bulk these quantities lead to the so-called magnetoelectric effect, in which the polarization is proportional to the magnetic field or the magnetization is proportional to the electric field. This effect can be exploited in order to obtain other exotic phenomena such as magnetic monopole charge-image \cite{qi}. In that region inside TI, $\theta$ is a constant of value equal to $\pi$, and thus the electrodynamics laws in material media become unchanged because the extra term is just a total derivative \cite{Essin}.

On the boundary of the TIs under consideration, the Maxwell's equations are modified besides the constitutive relations, because the $\theta$ axionlike factor presents a nonzero gradient, since in regions external to the TI (vacuum or ordinary insulators), $\theta=0$, and inside it, $\theta=\pi$. Thus, the Gauss's law changes to
\begin{equation}\label{3}
{\bf\nabla\cdot E}=4\pi\rho-\frac{\alpha c}{\pi}{\bf\nabla\theta\cdot B}.
\end{equation}

The electric field is necessary in our analysis for two reasons: Firstly, the theta term does not exist without it. Secondly, it induces the electric charge concentration on the conducting surface. In fact, the experiments with Bi$_2$Se$_3$ described in Ref. \cite{Tong} used magnetic fields as well as biased electric fields in order to study the Landau quantization on the TI boundary, whose energy levels must be taken into account when one calculates the vacuum energy \cite{Beneventano}. Then the total charge density is a combination of that one induced by the external electric field, $\rho_{ind}$, with the one of topological origin due to the theta term, $\tilde{\rho}=\rho_{top}$.

Thus, after the integration of Eq. (\ref{3}) throughout a thin pill-box-type (cylindrical) volume encompassing the TI surface and within the limit of the fields action, we get
\begin{equation}\label{4}
\tilde{q}=q_{total}-{q}_{ind}=\frac{\alpha c}{4\pi^2}B\int\int dxdy\int^{z_{out}}_{z_{in}} \frac{\partial \theta}{\partial z}dz=\frac{\alpha c}{4\pi}\Phi,
\end{equation}
where $\widetilde{q}$ is the effective (topological) charge and $\Phi$ is the magnetic flux on an area contained in the TI surface. Without the magnetic flux the induced charge is equal to the total one. This flux cannot be arbitrarily low, since the $\theta$-term in the Lagrangian (\ref{1}) is introduced after accounting for the proper quantization of the magnetic flux \cite{Hoyos}. If we take into account the quantum of magnetic flux, $\Phi_0=h/2e$, then $\widetilde{q}=e/4$. In this case we would be dealing with effective electric charges which are fractionary ones, carried by quasiparticles.

\section{Polarized Vacuum Energy on the TI Boundary}

It is known that dynamical effects arise from the variations in the vacuum expectation value of observables associated with a quantum field, imposed by external boundary conditions or by nontrivial topologies of the physical space as well as by interactions with an external field generated from a particular source. The difference between the vacuum energies of these configurations is named Casimir energy, in the two first cases, and polarized vacuum energy, in the last one. In this section, we will consider the configuration of a static and spatially homogeneous magnetic field acting perpendicularly on the surface of a $Z_2$-type TI with TR symmetry in order to calculate the polarized vacuum energy of a spin 1/2 field situated on this edge, in both the weak-field approximation and general case.

\subsection{Weak-field approximation}

In the weak-field regimen, the energy gap opening caused by the incident magnetic field is neglected by considering the dispersion relation associated with the surface states of a TI around the Dirac points, given by \cite{Qien}
\begin{equation}\label{5.0}
\epsilon({\bf p})\approx \pm{\bf v}_F\cdot {\bf p},
\end{equation}
where ${\bf v}_F$ is the Fermi velocity. Thus, we will deal with the polarized vacuum energy of a massless fermion in 2D space; the ordinary Lagrangian density of an electromagnetic field without sources plus a spinor field with charge $q$ interacting with the field is, in covariant language, given by
\begin{equation}\label{5}
\mathcal{L}_{qed}=-\frac{1}{4}F_{\mu\nu}F^{\mu\nu}+\overline{\Psi}\gamma^{\mu}(i\partial_{\mu}-qA_{\mu})\Psi,
\end{equation}
where $\gamma^{\mu}$ are the Dirac's $2 \times 2$ matrices (in 1+2 dimensions) and $\overline{\Psi}=\Psi^{\dag}\gamma^0$. The gauge choice is
\begin{equation}\label{5.a}
A_\mu\equiv(0,0,-\frac{1}{2}\rho B,0),
\end{equation}
in cylindrical coordinates $(\rho,\phi,z)$, compatible with a constant magnetic field $B$ pointing in the $z$ direction. The polarized vacuum energy density, $\epsilon_C$, related to the spin 1/2 field in 1+2 dimensions for this particular gauge and for massive fermions, was found in \cite{Visser} by different methods, and it is given by
\begin{equation}\label{6}
\epsilon_C=\frac{2qB}{2\pi}\left[\frac{m}{2}-\sqrt{2qB}\zeta\left(-\frac{1}{2},\frac{m^2}{2qB}\right)\right]-\frac{m^3}{3\pi},
\end{equation}
in natural units ($\hbar=c=1$), with $\zeta(s,q)$ being the Hurwitz zeta function \cite{Milton}, which is the ordinary Riemann zeta function when the second entry vanishes. The last term in the expression above is the vacuum energy density of the spinor field with the magnetic field switched off. It is worth pointing out that this polarized vacuum energy did not need to pass by the renormalization of some parameters, since it is naturally finite, at least in 1+2 dimensions. We also remark that such energy does not generate any physical influence because it does not depend on a geometrical parameter.

By considering now the surface of a $Z_2$-type TI with TR symmetry, we have seen that one must add the axion-like term to the Lagrangian (\ref{5}), given in covariant notation by
\begin{equation}\label{6a}
\frac{c\alpha}{4\pi^2}\theta\epsilon_{\mu\nu\rho\sigma}F^{\mu\nu}F^{\rho\sigma}=\frac{c\alpha }{4\pi^2}\theta{\bf E}\cdot{\bf B},
\end{equation}
where $\epsilon_{\mu\nu\rho\sigma}$ is the four-dimensional completely antisymmetric tensor. Therefore, the modified electromagnetic sector of the Lagrangian allows us to take into account the electric charge derived in Eq. (\ref{4}), since this charge will correct the vacuum energy found in the literature, given by Eq. (\ref{6}). Such a charge is responsible for the dynamical effect that we are looking for and it does not occur when ordinary charges alone are considered. Thus this topological charge must be taken into account in the coupling terms, and it will be assumed to possess fermionic nature because it is just a correction to the induced (ordinary) charges. Beside this, we must have $m\approx0$, provided that we are dealing with carriers on a graphenelike conducting sheet that forms the TI surface, in which the linear dispersion for massless fermions is obeyed, at least around the Dirac's points \cite{Castro}, as we have pointed out. We also must change $c\rightarrow v_F$, where this latter is the Fermi velocity. Joining all this and following \cite{Visser}, we arrive at
\begin{eqnarray}\label{7}
\epsilon_C&=&-\frac{2\widetilde{q}B}{2\pi}\left[\sqrt{2\widetilde{q}B}\zeta\left(-\frac{1}{2},0\right)\right]\nonumber\\
&\approx&\frac{0.207}{2\pi\hbar^{1/2}v_F^{1/2}}\left[2\left(\frac{\alpha v_F\Phi}{4\pi}\right)B\right]^{3/2},
\end{eqnarray}
for the same gauge (\ref{5.a}), after restoring the constants and $\zeta(-1/2,0)\approx -0.207$.

The above result can also take into account the fact that the considered surface forms an interface between two TIs. In this case, the more general expression for the axionlike parameter is $\theta=\pi(2n+1)$, $n=0,\pm1,\pm2...$ \cite{Marcel}, and the expression for the polarized vacuum energy density becomes
\begin{equation}
\epsilon_C\approx\frac{0.207}{2\pi\hbar^{1/2}v_F^{1/2}}\left\{2\left[\frac{(k-n)\alpha v_F\Phi}{2\pi}\right]B\right\}^{3/2},
\end{equation}
yielding a quantization of the polarized vacuum energy density. Notice that $k\in\mathbb{Z}$, too.

If we consider that the magnetic flux is given in a function of a linear dimension of the TI boundary, like $\Phi=B\pi R^2$, and if this flux is cylindrical with radius $R$, then the polarized vacuum energy density depends on this length taken on the TI surface, which yields a density of radial force $f_C(R)$ given by
\begin{eqnarray}\label{8}
f_C(R)=-\frac{d\epsilon_C}{dR}\approx-3\frac{0.207(2\alpha)^{3/2} v_FB^3R^2}{16\pi\hbar^{1/2}}.
\end{eqnarray}
This Casimir force density is tangential to the TI surface, and can be seen as a radial shear stress that arises due to the quantum vacuum influence. At first instance, through the signal of this force we deduce that it tends to shrink the surface.

\subsection{General case}

Because a sufficiently strong magnetic field  opens an interband gap on the boundary, Dirac fermions acquire an effective mass $m$, which can be expressed as a function of the external magnetic field in the form
\begin{equation}\label{9}
\Delta=mv_F^2=\widetilde{\mu}B,
\end{equation}
with $\widetilde{\mu}$ being the magnetic momentum of the carrier \cite{Shen}. Thus we must use Eq. (\ref{6}) and adopt the same procedure made in the previous section; {\it i.e.}, we must take into account the effective electric $\widetilde{q}$ charge in order to get
\begin{equation}\label{10}
\epsilon_C=\frac{2\widetilde{q}B}{2\pi}\left[\frac{\Delta}{2}-\sqrt{2\widetilde{q}B}\zeta\left(-\frac{1}{2},\frac{\Delta^2}{2\widetilde{q}B}\right)\right]-\frac{\Delta^3}{3\pi},
\end{equation}
where, once more, $\widetilde{q}=\frac{\alpha v_F}{4\pi}\Phi$. Explicitly, in terms of the geometric parameter $R$, we have that
\begin{equation}\label{10}
\epsilon_C=\frac{2kB^2R^2}{2\pi}\left[\frac{\widetilde{\mu}B}{2}-\sqrt{2k}\zeta\left(-\frac{1}{2},\frac{\widetilde{\mu}^2}{2kR^2}\right)BR\right]-\frac{(\widetilde{\mu}B)^3}{3\pi},
\end{equation}
where $k=\alpha v_F/4\pi$.

\begin{figure}[!h]
\includegraphics[scale=0.9]{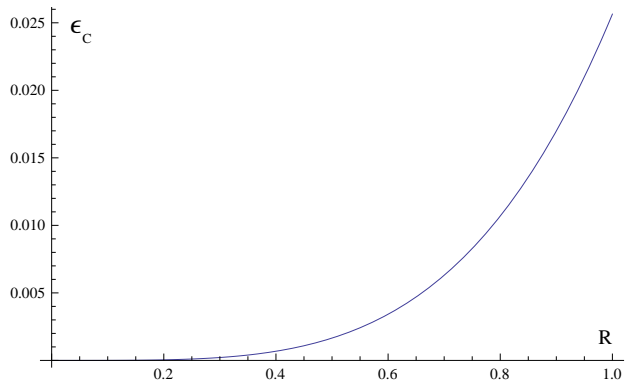}
\caption{Polarized vacuum energy as a function of the radius $R$, in unities such that the non-numeric parameters are made equal to 1.}
\end{figure}

In Fig. 1 we depict the behavior of the polarized vacuum energy (designed here as Casimir energy) as a function of the radius $R$ of the region in which the magnetic field acts ({\it i.e.}, the flux ``tube'' radius), taking all the non-numeric parameters equal to unity. We remark the monotonic energy increasing with this geometric parameter vanishes at $R=0$ for any $B$ and $\widetilde{u}$. For large values of $R$, the polarized vacuum energy increases almost quadratically with this latter. We can still notice the cubic dependence of that energy on the magnetic field, such that it also grows monotonically with the field, for $B>0$.

From Eq. (\ref{10}) we calculate the radial force, which is given by
\begin{eqnarray}
f_C&=&-\frac{2 B^2 k R}{\pi } \left[\frac{B \widetilde{u}}{2}-\sqrt{2} B \sqrt{k} R \zeta\left(-\frac{1}{2},\frac{\widetilde{u}^2}{2 k R^2}\right)\right]\nonumber\\
&-&\frac{B^2 k R^2}{\pi }\left[-\sqrt{2} B \sqrt{k} \zeta\left(-\frac{1}{2},\frac{\widetilde{u}^2}{2 k R^2}\right)\right]\nonumber\\
&-&\frac{B^2 k R^2}{\pi }\left[\frac{B \widetilde{u}^2}{\sqrt{2} \sqrt{k} R^2} \zeta\left(\frac{1}{2},\frac{\widetilde{u}^2}{2 k R^2}\right)\right].
\end{eqnarray}
It is possible to show that this force diminishes with the radius $R$ for any positive value of $B$ and $\widetilde{u}$, becoming always more negative, which yields a growing shear stress that tends to shrink the surface.\\

\section{Concluding Remarks}

In this paper we have investigated the polarized vacuum on the surface of a class of TIs, labeled by the topological index $Z_2$ and exhibiting time reversal (TR) symmetry, at zero temperature. Initially it was discussed how the Maxwell's laws with source terms are modified in such systems, leading to the correction of the induced electric charge distribution by virtue of the presence of an axionlike pseudoscalar factor and when one considers both uniform magnetic and electric fields perpendicular to the boundary under analysis. Since the electric field does not influence the Landau levels on the TI surface \cite{Tong} and consequently the respective polarized vacuum energy, it was only indirectly taken into account in the electric charge induced by the field.

Thus, we have obtained an effective charge which is a correction to the induced ordinary charges due to the topological properties of the system and calculated the modifications in the polarized vacuum energy of a spin 1/2 field confined on one of the TI conducting edges, in both the weak-field approximation and general case. For the first case, we have also obtained the polarized vacuum energy density for the situation in which the surface forms an interface between two TIs, and found an expression that points out the quantization of this energy.

We have noticed that, differently from the ordinary polarized vacuum energy of 2-D fermions presented in the literature, the result found by us depends on a length parameter defined on the boundary, which yields a tangential force (in fact, a radial shear stress) associated to it. The negative signal of the found force indicates a tendency for it to shrink the surface. It is worth pointing out that the ordinary charges induced by the electric field do not generate alone any dynamical effects because they do not carry information on the topological properties of the system and thus the polarized vacuum energy is independent of its geometrical parameters \cite{Visser,Beneventano}.

\section*{ACKNOWLEDGEMENTS}

M. O. Tahim and G. D. Saraiva would like to thank CNPq for partial financial support.
\newpage


\end{document}